\documentclass{article}
\usepackage[]{emulateapj}

\def\beq{\begin{equation}}
\def\eeq{\end{equation}}
\def\fun#1#2{\lower3.6pt\vbox{\baselineskip0pt\lineskip.9pt
  \ialign{$\mathsurround=0pt#1\hfil##\hfil$\crcr#2\crcr\sim\crcr}}}
\def\lap{\mathrel{\mathpalette\fun <}}
\def\gap{\mathrel{\mathpalette\fun >}}

\lefthead{Merritt and Cruz}
\righthead{Cusp Disruption}

\begin{document}

\title{Cusp Disruption in Minor Mergers}
\author{David Merritt$^1$ and Fidel Cruz$^{1,2}$}
\affil{$^1$Department of Physics and Astronomy, Rutgers University}
\affil{$^2$Instituto de Astronomia, UNAM, Ensenada, Mexico}

\begin{abstract}
We present $0.55\times 10^6$-particle simulations of the accretion 
of high-density dwarf galaxies by low-density giant galaxies, 
using models that contain both power-law central density cusps and
point masses representing supermassive black holes.
The cusp of the dwarf galaxy is disrupted during the merger,
producing a remnant with a central density that is only slightly 
higher than that of the giant galaxy initially.
Removing the black hole from the giant galaxy allows
the dwarf galaxy to remain intact and leads to a remnant with
a high central density, contrary to what is observed.
Our results support the hypothesis that the persistence of 
low-density cores in giant galaxies is a consequence of
supermassive black holes.

\end{abstract}


\section{Introduction}

Bright spheroids (elliptical galaxies and the bulges of spiral galaxies)
tend to be less dense than faint spheroids, in two distinct ways.
Bright spheroids have larger scale lengths, 
$r_e\sim L^{3/4}$, and lower mean densities within $r_e$,
where $r_e$ is the half-light radius and $L$ is the total luminosity.
Bright galaxies also tend to have density profiles that are less 
concentrated toward the center.
The luminosity densities of elliptical galaxies and bulges rise
approximately as power laws at the smallest observable radii,
$\rho_L\sim r^{-\gamma}$.
Faint galaxies ($M_{\rm V}\gap -20$) have $1.5\lap\gamma\lap 2.5$ 
while bright galaxies have $0\lap\gamma\lap 1.5$ 
(\cite{cra93}; \cite{fer94}; \cite{geb96}).

The existence of the second correlation raises two questions.
Why do bright galaxies have lower central concentrations than
faint galaxies; 
and how is this correlation maintained in the face of mergers?
Big galaxies will sometimes accrete small galaxies,
and the small galaxy will be resistant to tidal disruption 
due to its higher density (\cite{kor84}; \cite{baq89}).
A small galaxy that survived such a merger with its central regions
intact would create a new, high-density core in the merger remnant,
destroying the observed correlation between $L$ and $\gamma$.

The survivability of compact galaxies during mergers is relevant
to the origin of galaxies with kinematically distinct cores.
Forbes, Franx \& Illingworth (1995) investigated whether the properties
of such galaxies were consistent with the accretion hypothesis.
They found no strong evidence for abnormal surface brightness profiles
(`cores within cores') but noted that ``This problem can possibly
be circumvented by adding a massive black hole to the host galaxy.
The [host] galaxy might be able to disrupt the victim galaxy,
resulting in a density profile more like that which is seen.''
They suggested that ``Simulations with realistic density profiles
and possible black holes would be valuable.''

Here we present the results of such simulations.
We consider mergers between initially spherical galaxies with
a mass ratio of $10:1$; each galaxy has a power-law
central density cusp as well as a central point mass representing a
supermassive black hole.
The galaxies' structural parameters are scaled in accordance
with observed relations (\S 2).
We find, in agreement with the suggestion of Forbes, Franx \& Illingworth
(1995), that the black hole in the primary galaxy is effective at
disrupting the cusp of the secondary galaxy, 
yielding a remnant density profile
that is only slightly more centrally concentrated than that of the
primary galaxy at the start of the simulation.

\section{METHOD}

The initial galaxies were generated from Dehnen's (1993) spherical
density law, 
$\rho(r) = \left[(3-\gamma)M/4\pi a^3\right](r/a)^{-\gamma}(1+r/a)^{\gamma-4}$.
Initial particle velocities were assigned from the unique isotropic 
distribution function that reproduces Dehnen's density law in the
combined potential of the stars and a central point mass representing
the black hole (\cite{tre94}).
Our primary galaxy had $\gamma=1$, characteristic of the shallow
cusps of bright galaxies,
and our secondary galaxy had $\gamma=2$, 
characteristic of the steep cusps of dwarf galaxies.
Each ``black hole'' was assigned a mass of $2\times 10^{-3}$ times 
that of its parent galaxy.
This is slightly above, but consistent with,
the current best estimate of $\sim 0.0012$ for the 
average ratio of black hole mass to luminous galaxy mass in the 
local universe (\cite{mef01}).
Henceforth the subscripts $1$ and $2$ will refer to the primary (massive)
and secondary (dwarf) galaxies respectively.
All of the simulations had $M_1/M_2=10$.

The mass $M_1$ and scale length $a_1$ of the primary galaxy were set to 
unity.
The orbital period at the half-mass radius of the primary is $\sim 33.3$ 
in model units.
The scale length of the secondary galaxy was computed from the relation
\beq
{a_2\over a_1} = {r_{e,2}\over r_{e,1}}\left({r_{e,1}\over a_1}/{r_{e,2}\over a_2}\right)
\eeq
where the ratio $r_e/a$ between effective radius and scale length is 
given by Dehnen (1993) as a function of $\gamma$.
Real spheroids have $r_e\propto L^{\alpha}$ and $M/L\propto L^{\beta}$ with
$M/L$ the mass-to-light ratio of the stars; hence $r_e\propto M^{\alpha/(\beta+1)}$.
Setting $\alpha\approx 0.75$ (e.g. \cite{vam98}) and 
$\beta\approx 0.25$ (e.g. \cite{fab87}) gives
$r_{e,2}/r_{e,1}\approx (M_2/M_1)^{3/5}$.
Using our assumed mass ratio of $10:1$ and the adopted values for $\gamma$,
equation (1) then implies $a_2/a_1 = 0.618$.

The center of the secondary galaxy was initially displaced
from the center of the primary galaxy by 
$\sim 3$ times the half-mass radius of the 
primary along the $y$-axis.
The initial velocity of the secondary galaxy was assigned one
of four values determined by $\kappa\equiv L/L_{circ}$, 
the angular momentum of the secondary's orbit in units of the
angular momentum of a circular orbit.
We chose $\kappa=\left\{ 0, 0.2, 0.5, 0.8\right\}$; the initial
velocity vector of the secondary galaxy was parallel to the $x$ axis.

The primary galaxy was assigned $N_1=0.5\times 10^6$ particles and 
the secondary galaxy $N_2=0.05\times 10^6$ particles. 
Since $N_1/N_2=M_1/M_2=10$, all particles had the same mass.
This was done to avoid any spurious relaxation due to mass segregation.

The evolution was followed using the tree code {\tt GADGET}
of Springel et. al (2000), 
a parallel algorithm with continuosly variable time steps
for each particle. 
The parameters in this code that most strongly affect the 
accuracy of the integrations 
are the softening length $h$ and the time step parameter $\eta$.
We chose $h=8\times 10^{-4}$, slightly smaller than the
radius $\sim 10^{-3}$ at which the two black holes would be expected
to form a hard binary.
This softening length corresponds to $\lap 1$ pc with typical scalings.
Our value $\eta = 0.008$ for the accuracy parameter was chosen so as to 
require $\gap 10^2$ time steps for a star in a circular orbit with radius 
$h$ around the large black hole.
Extensive testing was carried out to ensure that the selected values
for the tree-code parameters resulted in no changes in the 
density profiles of either galaxy on length scales $\gap 10^{-3}$
when integrated in isolation.
The merger simulations were continued until roughly one crossing time
of the primary galaxy after formation of the hard black-hole
binary. 
Integrations were carried out on the parallel supercomputers at
the Rutgers Center for Advanced Information Processing 
(a Sun HPC-10000) and at the San Diego Supercomputer Center
(Cray T3E).
All simulations used $16$ processors.

In order to test the influence of the secondary's black hole on the 
evolution, each run was repeated using a secondary galaxy with the same 
initial density profile but lacking a central point mass.
To test the importance of the primary's black hole, 
a further set of runs were carried out in which neither
galaxy contained a black hole.
The velocity distribution functions for the models without
black holes were computed as in Dehnen (1993).

\section{RESULTS}

The results are summarized in Figures 1 and 2.
Figure 1a shows, for $\kappa=0.2$,
the number of particles initially associated with the two galaxies 
that remain within spheres of radius $0.1$ (primary) and $0.05$ 
(secondary) around their respective black holes;
these are roughly the respective radii inside of which the final
density profile of each galaxy shows substantial evolution.
Figure 1 reveals two ways in which the merger causes the central
density of the final object to be less than the sum of the 
central densities of the two initial objects.
1. The density of stars around the black hole in the primary galaxy
drops steadily throughout the simulation, 
and drops suddenly whenever the secondary galaxy passes near 
the primary's center; 
the drops are preceded by jumps resulting from the impulsive 
increase in the gravitational force caused by the passage of the small galaxy.
2. The density of stars around the black hole in the secondary galaxy
also drops, but only after the secondary passes sufficiently close
to the center of the primary.
In the case of plunging orbits, $\kappa=0$ or $0.2$, the first
close passage occurs during the first infall;
the drop in density of the secondary is again preceded by a jump indicating
that the mechanism is impulsive addition of energy,
this time from the gravitational force of the primary's black hole.
In the case of the more circular orbits, $\kappa=0.5$ and $0.8$, 
the central density of the secondary remains high until
the radius of its orbit has decayed to $\sim 0.1$.

The presence of the black hole in the secondary galaxy might naively
be expected to contribute to the stability of that galaxy's core 
by enhancing the gravitational force there.
In fact the opposite is true: the density
at late times in the runs without a second black hole is generally 
{\it higher} than in the run with two black holes
(Fig. 1a).
This is probably a consequence of three-body ejection of stars
by the pair of black holes (\cite{qui96}).
However the softening of the inter-particle forces in {\tt GADGET}
does not permit this process to be followed with high accuracy.
The final central densities that we observe in the runs with
two black holes should therefore be interpreted as upper limits.

Removing the black hole from the primary galaxy has almost
no effect on its evolution, but the central density of the secondary
galaxy now remains high throughout the merger (Figure 1b).

Figures 2a-d show the final density profiles from the four runs
in which both galaxies contained black holes.
The profiles are centered on the black hole binary and were
computed using {\tt MAPEL} (\cite{mer94}).
The secondary's density (Fig. 2a) is lowered at small radii,
$r\lap 0.05$, due to impulsive heating from the primary black hole,
and at large radii, $r\gap 0.1$, due to tidal forces from the
primary galaxy.
The greatest decrease in central density occurs in the runs with
lowest $\kappa$, i.e. the most plunging orbits.
In the case of stars initially associated with the primary galaxy,
the final density profile is remarkably independent of
$\kappa$ (Fig. 2b).
The primary galaxy was found to always exhibit a $\rho\sim r^{-0.5}$ 
central density cusp at the end of the simulations;
a similar result was found by Nakano \& Makino (1999a,b)
when they dropped a black hole into the core of a galaxy
containing no black hole.
The net result (Fig. 2c) is a remnant density profile that rises
modestly above that of the initial, primary galaxy at small radii,
more so in the runs with larger $\kappa$.
However the slope of the surface brightness profile 
(Fig. 2d) is only slightly increased near the center
compared with that of the primary galaxy.
We note that many bright elliptical galaxies also show
mild inflections in their central surface brightness
profiles on similar scales (e.g. Figure 1 of \cite{mef95}),
perhaps relics of past mergers.

When the black hole is removed from the primary galaxy, 
the secondary galaxy survives the merger intact, 
with almost no change in its density profile at small radii (Fig. 2e).
The final density profile of the remnant
is roughly a superposition of the two initial profiles
(Figure 2g);
a substantial inflection in the surface brightness profile
now appears at a radius of $\sim 0.1$ (Figure 2h).
Profiles of this sort are rarely if ever observed in the brightest 
elliptical galaxies (\cite{fer94}; \cite{geb96}).

\section{Discussion}
Our merger simulations are the first in which the central densities
of the merging galaxies are realistically high and in which both
galaxies contain ``black holes.''
Barnes (1999) described mergers of identical Dehnen models with
cusp slopes $\gamma=\left\{ 1, 1.5, 2\right\}$ but without 
black holes; 
he found that the cusps survived the mergers with little change.
We find a similar result here in the runs without black holes
except that the density of the primary galaxy falls due to
energy input from the secondary.
The opposite case -- mergers between galaxies with central
black holes but no cusps -- has been treated extensively
(e.g. \cite{emo91}; \cite{gcm94}; \cite{mak97}).
A number of authors have investigated the effects of dropping
one or two black holes into a pre-existing galaxy, either with
a central density cusp (\cite{quh97}) or without (\cite{nam99a}, 1999b).
The simulations closest in spirit to ours are those of
Holley-Bockelmann \& Richstone (2000).
These authors followed the evolution of a dense secondary galaxy
as it moved in the fixed gravitational field of a less-dense primary
containing a black hole; the secondary contained no black hole.
The decay of the secondary's orbit was induced artificially using
standard expressions for the dynamical friction, with corrections due
to the changing mass of the secondary.
Since the density of the primary galaxy was fixed in these simulations, 
the authors did not observe the drop in the primary's central density 
which we see here (Fig. 2b);
hence the central densities of their merger remnants were too high,
partly explaining their conclusion that all but the most plunging
orbits resulted in merger remnants with unphysically steep cusps.
Holley-Bockelmann \& Richstone 
also observed a stronger dependence of the secondary's
density evolution on its initial orbit than seen here, 
probably because of the approximate way in which orbital decay was
treated.

\begin{figure*}
\includegraphics{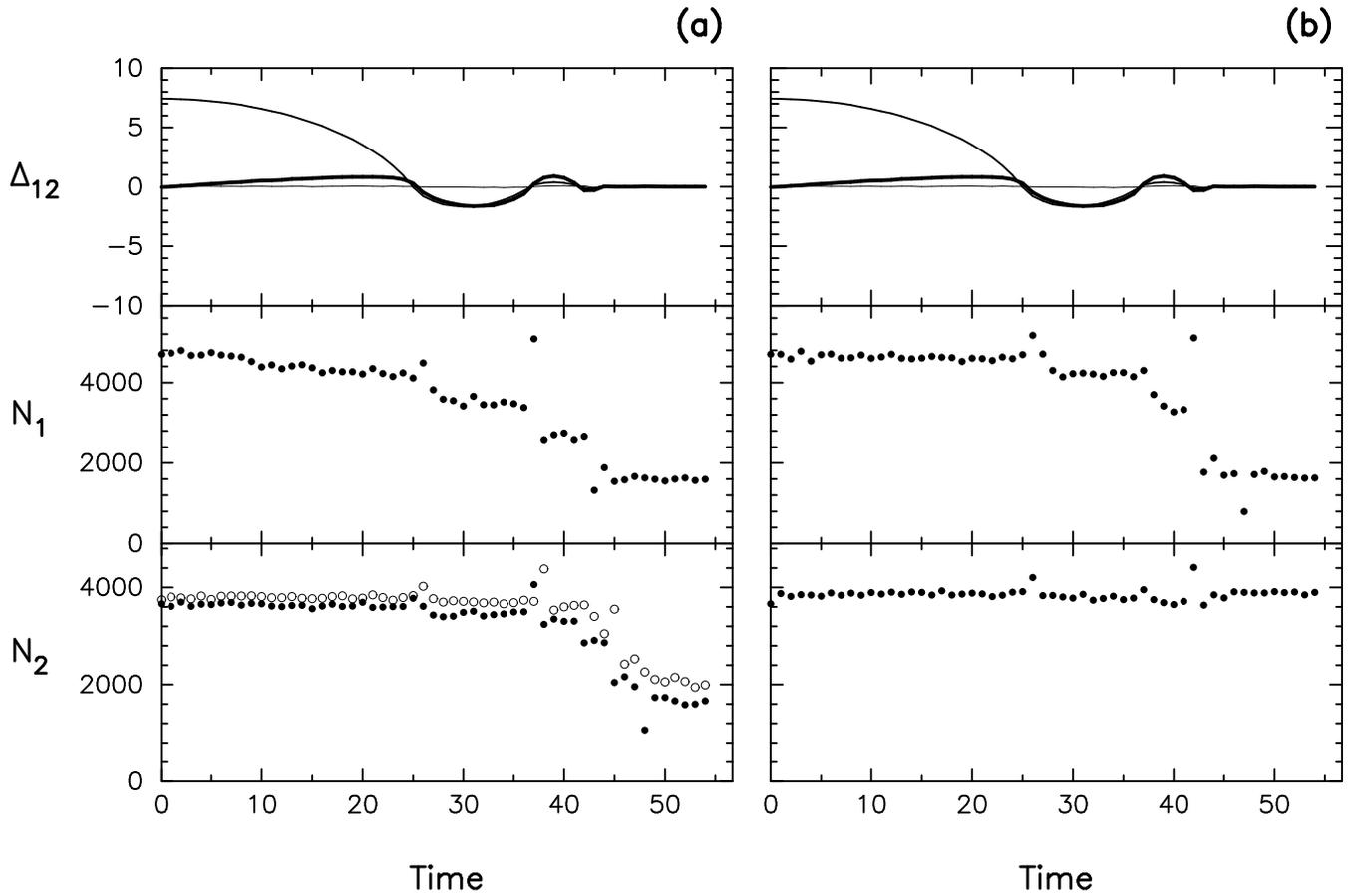}
\vspace{15.cm}
\caption{
Time evolution of the merger simulations with $\kappa=0.2$.
(a) Runs with a black hole in the primary galaxy.
Top panel: separation between the primary and secondary black holes,
in $x$ (thick line), $y$ (intermediate line), and $z$ (thin line).
Middle panel: number of particles initially associated with the primary
galaxy in a sphere of radius $0.1$ centered on its black hole.
Bottom panel: filled circles indicate number of particles initially
associated with the secondary galaxy in a sphere
of radius $0.05$ centered on its black hole.
Open circles correspond to runs in which the secondary galaxy contained
no black hole; $N_2$ is the number of particles in a sphere of radius
$0.05$ centered on the point of maximum density in the secondary galaxy.
(b) Similar to (a) except that neither galaxy contained a black hole.
Centers of both galaxies were defined as points of maximum density.
}
\end{figure*}

\begin{figure*}
\includegraphics{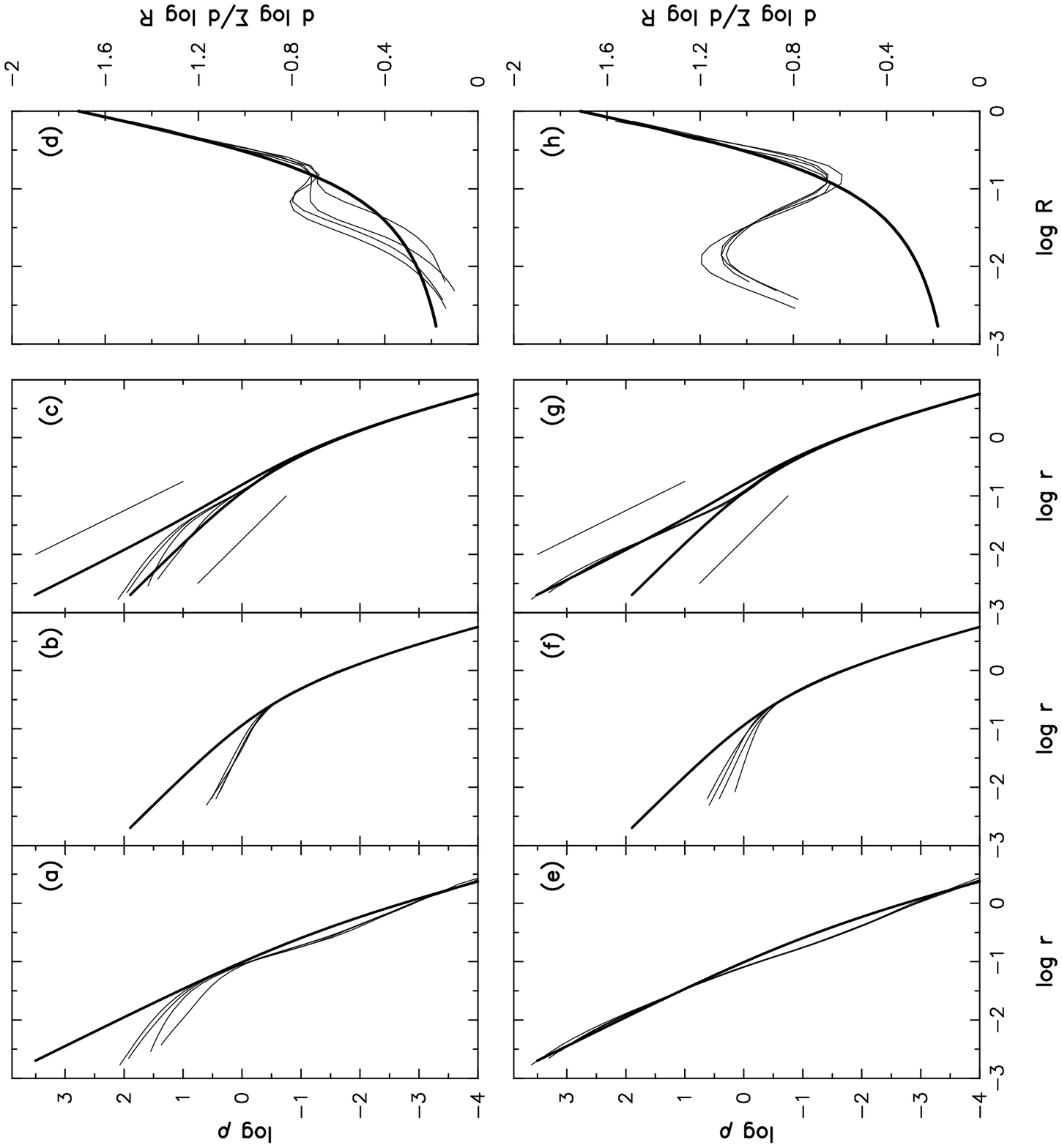}
\vspace{20.cm}
\caption{
Final density profiles from the simulations in
which each galaxy contained a black hole (a-d)
and in which neither galaxy contained a black hole (e-g).
The four thin curves in each frame correspond to the four different
orbits;
the thin curve extending farthest to the left corresponds
to $\kappa=0.8$.
(a), (e) Space density of stars initially associated with the secondary galaxy;
thick curves are the initial density profile.
(b), (f) Space density of stars initially associated with the primary galaxy;
thick curves are the initial density profile.
(c), (g) Space density of all stars.
Lower thick curves are the initial density profile of the primary galaxy,
and upper thick curves are the superposition of the initial density profiles
of the primary and secondary galaxies.
Lines of logarithmic slope $-1$ and $-2$ are also shown.
(d), (h) Logarithmic slope of the surface density profiles of the merger
remnants.
Thick curves correspond to the initial primary galaxy.
}
\end{figure*}

Our results support the hypothesis that the persistence of 
low-density cores in giant galaxies is a consequence of
the existence of supermassive black holes (\cite{ffi95}).
The central density did nevertheless increase in our simulations, 
and since a typical bright galaxy is expected to have accreted many
smaller galaxies since its formation (e.g. \cite{lac93}),
one might still predict the formation of dense nuclei in
bright galaxies, contrary to what is observed.
However our simulations are not able to follow the evolution of a 
black-hole binary on sub-parsec scales as it ejects stars from the
core and lowers the density there still more (\cite{qui96}).
Future work should be directed toward understanding the effects of
hierarchical mergers on galaxy density profiles, 
using $N$-body codes that can deal efficiently with unsoftened particles
and hence follow the interaction of a binary black hole with
the surrounding stars.

We thank V. Springel for advice about using {\tt GADGET} and
for making the code available to us prior to its public release.
The work described here was supported by NSF grants AST 96-17088 and 00-71099,
NASA grants NAG5-6037 and NAG5-9046, and by a fellowship from
the Consejo Nacional de Ciencia y Tecnologia de Mexico.
This work was partially supported by the National Computational 
Science Alliance under grant no. MCA00N010N and utilized the 
San Diego Supercomputer Center Cray T3E.
We are also grateful to the Center for Advanced Information Processing
at Rutgers University for their generous allocation of computer time.


\begin{thebibliography}{}

\bibitem[Balcells \& Quinn 1989]{baq89} 
	Balcells, M. \& Quinn, P. J. 1989, Ap. Space Sci., 156, 133

\bibitem[Barnes 1999]{bar99}
	Barnes, J. E. 1999, in Proceedings of IAU Symposium 186,
	Galaxy Interactions at Low and High Redshift, ed. J. E. Barnes
	\& D. B. Sanders, 137

\bibitem[Crane et al. 1993]{cra93}
	Crane, P. et al. 1993, \aj, 106, 1371

\bibitem[Dehnen 1993]{deh93}
	Dehnen, W. 1993, \mnras, 265, 250

\bibitem[Ebisuzaki, Makino \& Okumura 1991]{emo91}
	Ebisuzaki, T., Makino, J. \& Okumura, S. K. 1991, Nature, 354, 212

\bibitem[Faber et al. 1987]{fab87}
	Faber, S. M. et al. 1987, in Nearly Normal Galaxies, ed.
	S. M. Faber (Berlin: Springer), 175

\bibitem[Ferrarese et al. 1994]{fer94}
	Ferrarese, L., van den Bosch, F. C., Ford, H. C., Jaffe, W. \&
	O'Connell, R. W. 1994, \aj, 108, 1598

\bibitem[Forbes, Franx \& Illingworth 1995]{ffi95}
	Forbes, D. A., Franx, M. \& Illingworth, G. D. 1995, \aj, 109, 1988

\bibitem[Gebhardt et al. 1996]{geb96}
	Gebhardt, K. et al. 1996, \aj, 112, 105

\bibitem[Governato, Colpi \& Maraschi 1994]{gcm94}
	Governato, F., Colpi, M. \& Maraschi, L. 1994, \mnras, 271, 317

\bibitem[Holley-Bockelmann \& Richstone 2000]{hbr00}
	Holley-Bockelmann, K. \& Richstone, D. O. 2000, \apj, 531, 232

\bibitem[Kormendy 1984]{kor84}
	Kormendy, J. 1984, \apj, 287, 577

\bibitem[Lacey \& Cole 1993]{lac93} Lacey, C. \& Cole, S. 1993,
	\mnras, 262, 627

\bibitem[Makino 1997]{mak97}
	Makino, J. 1997, \apj, 478, 58

\bibitem[Merritt 1994]{mer94}
	Merritt, D. 1994, 
	http://www.physics.rutgers.edu/$\sim$merritt/mapel\_1.html

\bibitem[Merritt \& Ferrarese 2001]{mef01}
	Merritt, D. \& Ferrarese, L. 2001, \mnras, 320, L30

\bibitem[Merritt \& Fridman 1995]{mef95}
	Merritt, D. \& Fridman, T. 1995, in ``Fresh Views
	of Elliptical Galaxies,'' A.S.P. Conf. Ser. Vol. 86,
	ed. A. Buzzoni, A. Renzini \& A. Serrano, 13

\bibitem[Nakano \& Makino 1999a]{nam99a}
	Nakano, T. \& Makino, J. 1999a, \apj, 510, 155

\bibitem[Nakano \& Makino 1999b]{nam99b}
	Nakano, T. \& Makino, J. 1999b, \apj, 525, L77

\bibitem[Quinlan 1996]{qui96}
	Quinlan, G. 1996, New Astronomy, 1, 35

\bibitem[Quinlan \& Hernquist 1997]{quh97}
	Quinlan, G. \& Hernquist, L. 1997, New Astronomy, 2, 533 

\bibitem[Springel, Yoshida \& White 2000]{syw00}
	Springel, V., Yoshida, N. \& White, S. D. M. 2000, astro-ph/0003162

\bibitem[Tremaine et al. 1994]{tre94}
	Tremaine, S. et al. 1994, \aj, 107, 634

\bibitem[Valluri \& Merritt 1998]{vam98}
	Valluri, M. \& Merritt, D. 1998, \apj, 506, 686

\end{thebibliography}
\end{document}